\documentclass[preprint,amsmath,amssymb]{revtex4}

\usepackage{graphicx}
\usepackage{amsmath}
\usepackage{amsfonts}
\usepackage{bm}
\def\bfgrk #1{\mbox{\boldmath$#1$}}

\begin{document}

\title{Numerical Search for Periodic Solutions in the Vicinity of the Figure-Eight
Orbit: Slaloming around Singularities on the Shape Sphere}

%\titlerunning{Short form of title} % if too long for running head

\author{Milovan \v{S}uvakov\\
           Institute of Physics Belgrade, University of Belgrade, Pregrevica 118, 11080 Beograd, Serbia
           \email{suki@ipb.ac.rs} }

\begin{abstract}
We present the results of a numerical search for periodic orbits with
zero angular momentum in the Newtonian planar three-body problem with
equal masses focused on a narrow search window 
bracketing the figure-eight initial conditions.
We found eleven solutions that can be described as some power of the
``figure-eight'' solution in the sense of the topological classification
method.
One of these solutions, with the seventh power of the
``figure-eight'', is a choreography. We
show numerical evidence of its stability.


\end{abstract}

\maketitle

\section{Introduction}
\label{intro}

The search for periodic orbits in the three-body problem has a long history 
\cite{Poincare1899,Broucke1975,Hadjidemetriou1975a,Hadjidemetriou1975b,Hadjidemetriou1975c,Broucke1975b,Henon1976,Henon1977,Vanderbei:2004,Titov2012}. 
This work is follow-up of the previous study \cite{Suvakov2013} where a systematic numerical search 
for periodic solutions was started and some 13 new families of solutions were reported. Here we focus 
on a particular window in the space of initial velocities bracketing the figure-eight initial conditions 
(the initial positions are fixed at the so-called Euler point).
All eleven periodic solutions that we report here belong to (special) classes that can be 
described topologically as some power of figure-eight topological class.
One of these solutions, one with the seventh power of the ``figure-eight'', is a choreography.

Choreographies are a special kind of the three-body periodic orbits such that all bodies 
travel along the same trajectory in the plane, chasing one another around the trajectory at equal
time intervals. 
The first and simplest choreography ever found is the Lagrange solution (1772), in which three 
equal-mass particles move on a circle, while forming an equilateral triangle. That
solution, however, is both unstable and has non-vanishing angular momentum. The first stable
three-body choreographic orbit without angular momentum was found by Moore in 1993
\cite{Moore1993}. A formal variational existence proof for such solution is
given by Chenciner and Montgomery \cite{Chenciner2000}.

A large number (345) of three-body choreographies with non-zero angular momentum has been
found by Sim\'{o} \cite{Simo2002}, but they are all highly unstable \cite{Simo2012pc}. 
In papers \cite{Simo2002,Chenciner2002}, Sim\'{o} and co-authors have shown several solutions
including one choreography that they called "satellites of the eight". 
These satellite solutions, topologically speaking, might be associated with some power of figure-eight
orbit. Here we report the discovery of one new stable, zero-angular-momentum choreographic solution 
to the planar Newtonian three-body problem with equal masses. 

\begin{figure}[h!]
\centerline{\includegraphics[width=0.95\columnwidth,,keepaspectratio]{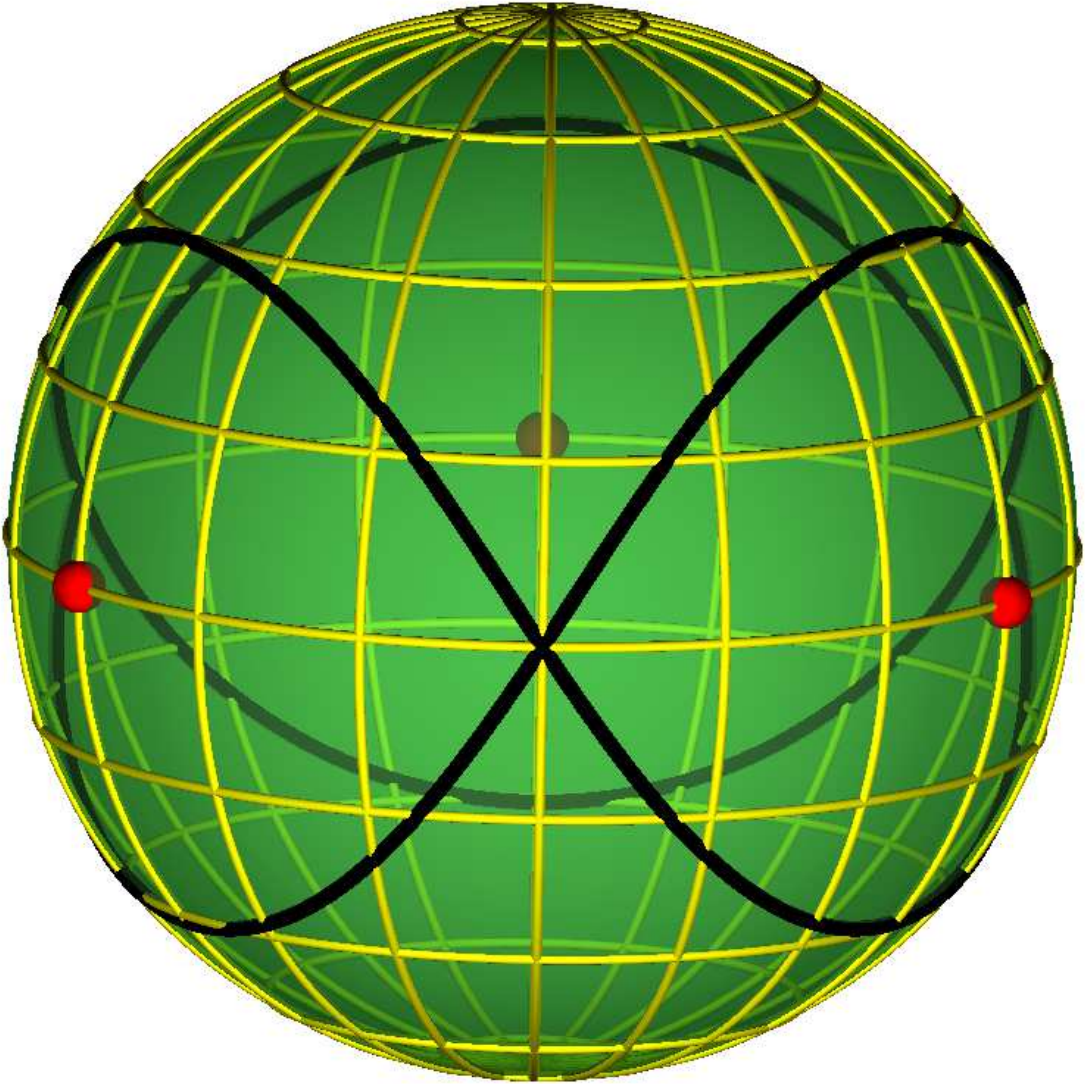}}
%\centerline{\includegraphics[width=0.95\columnwidth,,keepaspectratio]{grid8.eps}}
\caption{Figure eight orbit (black) on the shape-space sphere. Three two-body collision 
points (bold red) - singularities of the potential - lie on the ``equator''. } 
\label{fSFIG8}
\end{figure}

\section{Definitions and methods}
\label{defandm}

Before describing these solutions, we briefly remind the reader about the shape sphere coordinates 
that we are using for the classification of periodic solutions. 

\subsection{Shape sphere}
\label{ss}

With two three-body Jacobi relative coordinate vectors,
${\bfgrk \rho} = \frac{1}{\sqrt{2}}({\bf r_1} - {\bf r_2})$, 
${\bfgrk \lambda} = \frac{1}{\sqrt{6}}({\bf r_1} + {\bf r_2} - 
2 {\bf r_3})$, there are three independent scalar three-body variables.
The hyper-radius $R = \sqrt{\rho^{2} + \lambda^{2}}$ 
defines the ``overall size'' of the system and removes one of 
the three linear combinations of scalar variables. Thus,
one may relate the three scalars to a unit three-vector ${\hat {\bf n}}$ with the 
Cartesian components 
${\hat {\bf n}} = \left ( \frac{2 {\bm \rho} \cdot {\bm \lambda}}{R^2}, 
\frac{\lambda^2 - \rho^2}{R^2}, \frac{2 ({\bm \rho} \times {\bm \lambda}) 
\cdot {\bm e}_z}{R^2} \right )$. The domain of these three-body variables is a 
sphere with unit radius \cite{Iwai1987a}, \cite{Montgomery1993},
see Figure \ref{fSFIG8}. The equatorial circle corresponds
to collinear configurations (degenerate triangles) and the 
three points on it, Figure \ref{fSFIG8} correspond to two-body-collision singularities in the potential. 
The shape sphere together with the hyper-radius defines the configuration space of the planar 
three-body problem (the ``missing'' total rotation angle can be reconstructed from
the trajectory in this space and the condition of angular momentum conservation).

The ``figure-eight'' solution can be described as a slalom, i.e. as moving in a
zig-zag manner 
on the shape sphere, between the two-body singularities, while drifting
in the same general direction along the equator, e.g. eastward, or westward, see Figure \ref{fSFIG8}. 
As a consequence of parity, the number of full turns (the ``winding number'') around the shape 
sphere sufficient to reach the initial conditions must be even, and the minimal number is two, 
which is the case for the figure eight orbit. 

There is at least one other known solution, that makes two full turns around the shape
sphere, that is stable, but not a choreography: That is Sim\'{o}'s figure eight orbit (labeled by H3 in Ref.
\cite{Simo2002}). 
The question remains: are there any other periodic orbits with a higher
winding number? Such an orbit would have to ``miss'' the initial point in the phase space at each turn before 
the last one. Again, due parity, the winding number of such a trajectory has to be an even number $2k$, 
$k \in \mathbb{N}$.
If there are such solutions, then another question is: are there any choreographies among them? 

\begin{figure}[h!]
\centerline{\includegraphics[width=3.5in,,]{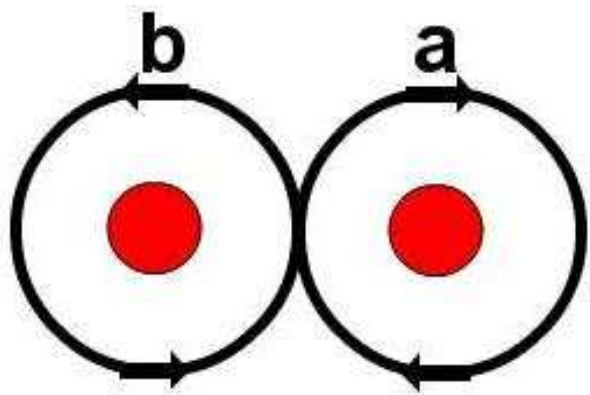}}
\caption{Diagrammatic representation of the two free group elements.} 
\label{f:freegroup}
\end{figure}

\subsection{Topological classification}
\label{topclass}

We use the topological classification of periodic three-body solutions, suggested by 
Montgomery \cite{Montgomery1998}. Considering a periodic orbit as a closed curve on a sphere 
with three punctures (the shape sphere with three singularities of the potential), the clasification 
of closed curves is given by the conjugacy classes of the fundamental group, which is, in this case, 
the free group on two letters $(a,b)$.

Graphically, this amounts to classifying closed curves according to 
their ``topologies'' on a sphere with three punctures. A stereographic projection of
this sphere onto a plane, using one of the punctures as the ``north pole'' 
effectively removes that puncture to infinity, and reduces the problem to one 
of classifying closed curves in a plane with two punctures. That leads to the 
aforementioned free group on two letters $(a,b)$, where (for definiteness) 
$a$ denotes a clockwise ``full turn/circle'' around the 
right-hand-side puncture, and $b$ denotes the counter-clockwise full turn/circle 
around the other pole/hole/puncture in the plane/sphere, see Fig. \ref{f:freegroup}. 

For better legibility we denote their inverses by capitalized letters $a^{-1}=A, b^{-1}=B$. 
Each family of orbits is associated with the conjugacy class of a free group element. 
For example the conjugacy class of $aB$ contains 
$A(aB)a = Ba$. The ``figure-eight'' orbit is related to the conjugacy class of the element $abAB$. 
The aforementioned curves with a ``winding number'' $2k$ belong to the conjugacy class of 
the group element $(abAB)^k$. 
Orbits of this kind will be called {\bf slaloms} and the exponent $k$ will be called 
the {\bf slalom power}.

\subsection{Search strategy}
\label{sforper}

Here we use the same numerical approach as in Ref. \cite{Suvakov2013}. The return proximity function in 
the phase space is defined as the absolute minimum of the distance from the initial condition by
\begin{equation}
d({\bf X_0},T_0)=\min_{t \le T_0} \left \vert {\bf X}(t)-{\bf X_0}\right \vert,
\end{equation}
where 
${\bf X}(t)=\left ({\bf r_1}(t),{\bf r_2}(t),{\bf r_3}(t),{\bf p_1}(t),{\bf p_2}(t),{\bf p_3}(t) 
\right )$ is a 12-vector in the phase space (all three bodies' Cartesian coordinates and 
velocities, i.e. without removing the center-of-mass motion), and ${\bf X_0}={\bf X}(0)$ 
is 12-vector describing the initial condition. 
We also define the return time $\tau({\bf X_0},T_0)$ as the time for which 
this minimum is reached. 
%\footnote{More formally $\tau({\bf X_0},T_0)=\argmin_{t \le T_0} \left \vert {\bf X}(t)-{\bf X_0}\right \vert$. Here $\vert ... \vert$ 
%presents the Euclidean norm, 
%though, for other purposes (e.g. searching for the reduced periodic
%solution), other metrics can be used.} 
Searching for periodic solutions with a period $T$ smaller then the parameter $T_0$ is clearly 
equivalent to finding zeros of the return proximity function.

The initial conditions for both Moore's \cite{Moore1993} and Sim\'{o}'s \cite{Simo2002} 
figure eight solutions can be found in the two dimensional subspace
of the eight-dimensional three-body phase space with the center-of-mass motion 
removed, see Fig. \ref{f:BIG}. 
Formally this two dimensional search plane is defined as the set of collinear congurations 
(``syzygies'') with one body exactly in the middle between the other two (the so-called Euler points), 
and with vanishing initial time derivative of the hyper-radius ${\dot R}=0$ and vanishing angular
momentum. In this subspace, all three particles' initial conditions can be specified by 
two parameters, the initial $x$ and $y$ components, $\dot x_1(0)$ and $\dot y_1(0)$, respectively, 
of a single velocity two-vector, as follows,  
$x_1(0)=-x_2(0)=-1$, $x_3(0)=0$, $y_1(0)=y_2(0)=y_3(0)=0$,  
$\dot x_2(0)=\dot x_1(0)$, $\dot x_3(0)=-2\dot x_1(0)$, $\dot y_2(0)=\dot y_1(0)$,
$\dot y_3(0)=-2\dot y_1(0)$.  

The differential equations of motion were solved numerically using the Runge-Kutta-Fehlberg 4.5 method 
and the return proximity function was calculated using linear interpolation between calculated 
trajectory points in the phase space. In all our calculations, the particle masses 
$m_1$, $m_2$, $m_3$ and Newton's gravitational constant $G$ were set equal to unity.

\section{Results}
\label{results}

\begin{figure}[tbp]
\centerline{\includegraphics[width=\columnwidth,,keepaspectratio]{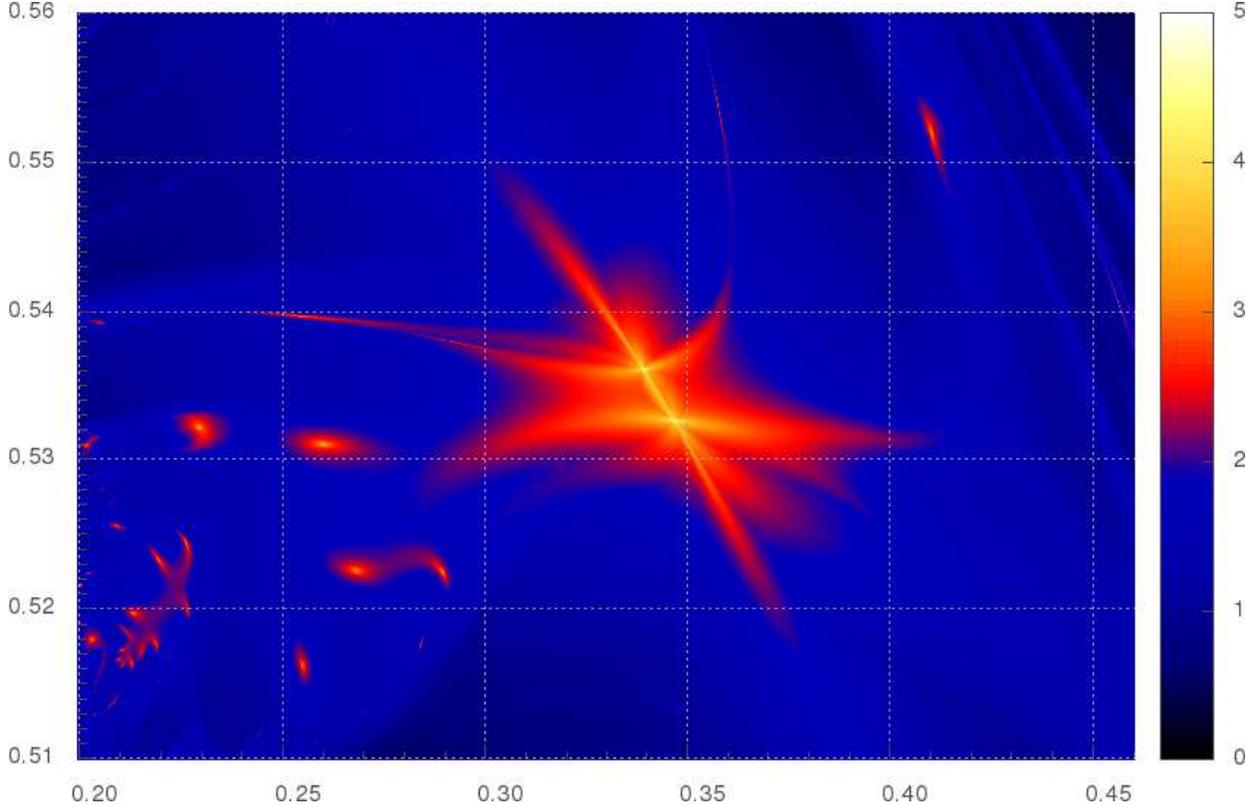}}
\caption{The decimal logarithm of the reciprocal of the return proximity function 
$-\log_{10} d({\bf X_0},T_0)$ in the search window around the initial conditions for the 
figure-eight solutions in search plane. 
On x-axis are the values of the initial velocity 
$\dot x_1(0) \in (0.20,0.46)$, and on the y-axis are the values of the initial velocity
$\dot y_1(0) \in (0.51,0.56)$.}
\label{f:BIG}
\end{figure}

We focused our numerical search on the (two-dimensional) search window in the
two dimensional search plane (defined above) around the
``figure-eight'' initial conditions: $\dot x_1(0) \in (0.20,0.46)$, $\dot y_1(0) \in
(0.51,0.56)$, see Figure \ref{f:BIG}. 
The equations of motion were integrated up to time $T_0=100$ for each initial condition out of 
$130 \times 1000$ possibilities (points on the grid) within the search window. 
The return proximity function $d({\bf X_0},T_0)$ was calculated and is shown in Figure \ref{f:BIG}. 
For each local minimum of the return proximity function lower then $10^{-4}$ 
(bright dots in Figure \ref{f:BIG}) on this grid we used the simple 
gradient descent algorithm to find the position of the minimum (root) more accurately. All 
minima below $10^{-6}$ are listed in Table \ref{tab:1} and can be seen in Figure \ref{f:BIG}. 
The initial conditions for Moore's ``figure-eight'' choreography and
Sim\'{o}'s ``figure-eight'' 
orbit are labeled by $M8$ and $S8$, respectively in Table I. All other labeled orbits
in Table I are slaloms of power $k$.

% The initial condition table
%=============================
\begin{table}[tbh]
\begin{center} 
\caption{Initial conditions and periods of three-body orbits.
${\dot x}_{1}(0), {\dot y}_{1}(0)$ are the first particle's initial 
velocities in the x- and y-directions, respectively, $T$ is the period, $k$ is slalom power 
(i.e. $abAB^k$ is homotopy class of the orbit), and the last column is the geometric-algebric (g-a) 
class (for explanation see text). We also list Moore's ($M8$) and Sim\'{o}'s ($S8$) figure-eight 
orbits, for comparison.}
\begin{tabular}{lc@{\hskip 0.1in}c@{\hskip 0.1in}c@{\hskip 0.1in}c@{\hskip
0.1in}c@{\hskip
0.1in}c@{\hskip
0.1in}c}
\hline \hline 
\setlength
 Label & ${\dot x}_{1}(0)$ & ${\dot y}_{1}(0)$ & ${\rm T}$ &  ${k}$ & g-a class\\
\hline
\hline
$M8$   & 0.3471128135672417 & 0.532726851767674 & 6.3250 & 1 & I.A \\ 
$S8$   & 0.3393928985595663 & 0.536191205596924 & 6.2917 & 1 & I.A \\ 
\hline
$NC1$ & 0.2554309326049807 & 0.516385834327506 & 35.042 & 7 & II.A \\
$NC2$ & 0.4103549868164067 & 0.551985438720704 & 57.544 & 7 & II.A \\
$O1$ & 0.2034916865234370 & 0.5181128588867190 & 32.850 & 7 & IV.A \\
$O2$ & 0.4568108129224680 & 0.5403305086130216 & 64.834 & 7 & IV.A \\
$O3$ & 0.2022171409759519 & 0.5311040339355467 & 53.621 & 11 & IV.A \\
$O4$ & 0.2712627822083244 & 0.5132559436920279 & 55.915 & 11 & IV.A \\
$O5$ & 0.2300043496704103 & 0.5323028446350102 & 71.011 & 14 & IV.A \\
$O6$ & 0.2108318037109371 & 0.5174100244140625 & 80.323 & 17 & IV.A \\
$O7$ & 0.2132731670875545 & 0.5165434524230961 & 80.356 & 17 & IV.A \\
$O8$ & 0.2138543002929687 & 0.5198665707397461 & 81.217 & 17 & III.A \\
$O9$ & 0.2193730914764402 & 0.5177814195442197 & 81.271 & 17 & III.A \\
$O10$ & 0.2272123532714848 & 0.5200484344272606 & 82.671 & 17 & IV.A \\
$O11$ & 0.2199766127929685 & 0.5234338500976567 & 82.743 & 17 & IV.A \\
$O12$ & 0.2266987607727048 & 0.5246235168190009 & 83.786 & 17 & III.A \\
$O13$ & 0.2686383642458915 & 0.5227270888731481 & 88.674 & 17 & III.A \\
$O14$ & 0.2605047016601568 & 0.5311685141601564 & 89.941 & 17 & IV.A \\
$O15$ & 0.2899041109619139 & 0.5226240653076171 & 91.982 & 17 & IV.A \\
\hline
\end{tabular}
\label{tab:1}
\end{center}
\end{table}
%===============================================================

There is a restriction on the slalom power $k$ in the case of choreographic solutions. 
If a solution is a choreography, the three masses folow each other, with a time delay of 
$T/3$, where $T$ is the full period. If one follows a choreography starting with a delay 
of $T/3$, one sees the same motion as without that delay, up to a cyclic permutation of 
three particles
\footnote{where the first mass is substituted into the position of the second, the second into 
the third, and the third into the position of the first.}. Algebraically, this
can be written as 
${\bf X}(t+T/3) = \hat{P} {\bf X}(t)$, where $\hat{P}$ is a cyclic permutation. 
The cyclic permutation $\hat P$ has a simple representation on the shape
sphere, {\it viz.} 
a rotation by $2\pi /3 $ around the vertical z-axis. For a choreography the azimuthal angle 
on the shape sphere after motion through time $T/3$ will be $2\pi k/3$. Therefore, 
the slalom power $k$ of any choreography can not be a number divisible by $3$ 
(otherwise $2\pi k/3$ is an integer multiple of $2\pi$, and the net rotation
mustnot be $2\pi /3$).

\begin{figure}[tbp]
\centerline{\includegraphics[width=0.95\columnwidth,,keepaspectratio]{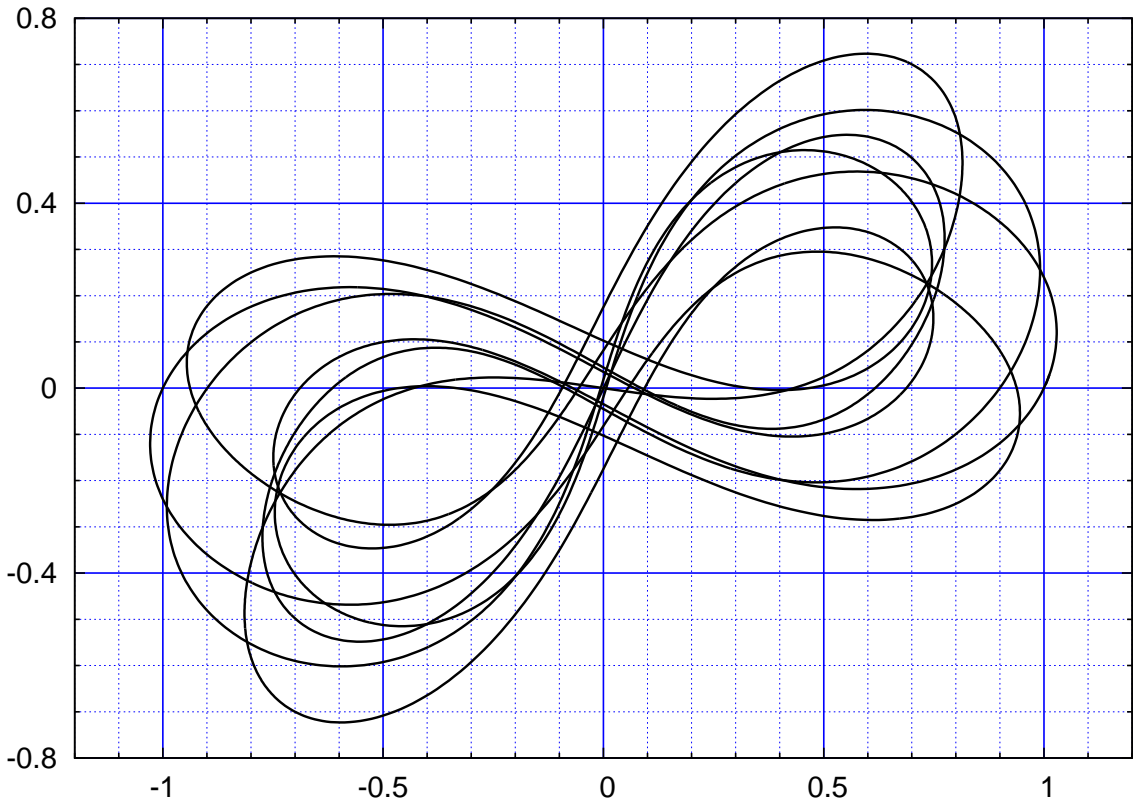}}
\caption{The trajectory of the new choreography in the (real) x-y plane.} 
\label{f:87}
\end{figure}

\subsection{The $k=7$ choreography}
\label{choreo}

All solutions listed in Table \ref{tab:1} satisfy the above choreography condition, 
but only solutions labeled $NC1$ and $NC2$ are choreographies. It turns out that these 
two solutions are equivalent, up to rescaling of temporal and spatial coordinates by
the scaling law \cite{Landau}: 
${\bf r} \rightarrow {\alpha}{\bf r}$, $t \rightarrow \alpha^{3/2} t$, and consequently 
${\bf v} \rightarrow {\bf v} / \sqrt{\alpha}$. The trajectory $HC1$ in real space is shown in 
Figure \ref{f:87}. 
This solution is centrally symmetric with respect to reflections
about the origin, which is the initial position of the ``middle'' mass, as a 
%straightforward 
consequence of the fact that all bodies follow the same trajectory,
that equations of motion are symmetric under time reversal and that 
all initial conditions are symmetric under the combined action of parity ${\cal P}$, 
time-reversal ${\cal T}$ and transposition $P_{12}$. 
It appears that our choreography does not have any additional symmetries in real space. 
The slalom power $k$ of this solution is seven. 

The trajectory in real space is composed of seven concatenated distorted figure-eight curves, 
Fig. \ref{f:87}. 
The trajectory passes through the coordinate origin twice. By following the position of the second 
mass one 
can see that the first passage through the origin corresponds to the initial time, whereas the 
second one corresponds to the point in time when the phase space position scales into the $NC2$ initial 
condition. 
The curves $HC1$ and $HC2$ in the $x-y$ plane are connected by a homothetic transformation 
with a homothety factor $\lambda \approx -1.3919$ and rotation angle of $\approx 0.252 \pi$ 
radians. The minus sign in the homothety factor $\lambda$ means that the two trajectories have 
opposite orientations, or in physical terms, that these two motions are time reversed.

\begin{figure}[tbp]
\centerline{\includegraphics[width=0.99\columnwidth,,keepaspectratio]{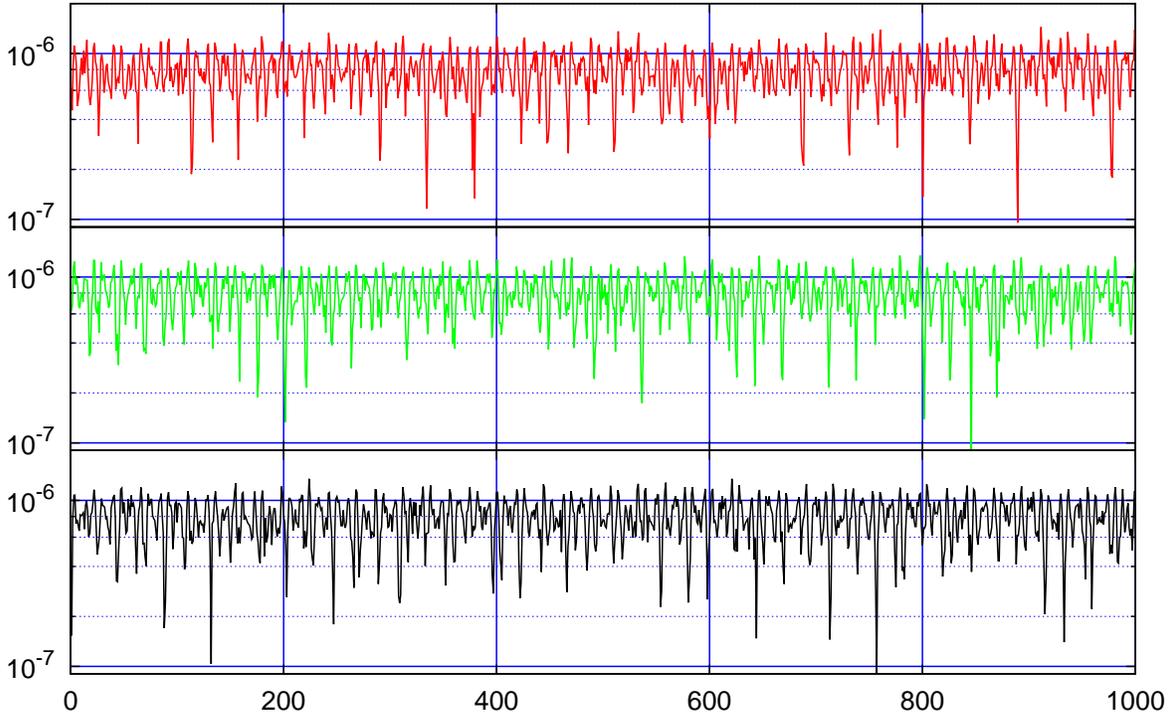}}
\caption{The minimal distance between initial condition and trajectory per each period $D_{0i}^T$ 
(bottom panel); and the minimal distance between initial coditions and two
cyclic permutations of 
the phase space coordinates: $D_{1i}^T$ and $D_{2i}^T$ (middle and top panel).} 
\label{fFIG87s}
\end{figure}

The key feature of this choreography is its stability. The return proximity per period: 
$D_{0i}^T=\min_{iT \le t < (i+1)T} \left \vert {\bf X}(t)-{\bf X_0}\right \vert$, where $T$ is 
period, is calculated up to a thousand periods (i.e. 42000 syzigies) and is shown in Fig \ref{fFIG87s}. 
In order to check if this solution is a true choreography, we show 
on the same plot: 
$D_{1i}^T=\min_{iT \le t < (i+1)T} \left \vert \hat{P}{\bf X}(t)-{\bf X_0}\right \vert,$ and 
$D_{2i}^T=\min_{iT \le t < (i+1)T} \left \vert \hat{P}^2 {\bf X}(t)-{\bf X_0}\right
\vert$, the minimal distances per 
period between the initial condition point and the two cyclic permutations of the trajectory coordinates 
in the phase space, respectively. 
Whereas the first array is zero for any periodic solution, all three arrays are zero when the solution 
is a choreography. One can see in Figure \ref{fFIG87s} that all three values fluctuate around an 
approximately constant level of {$10^{-6}$}. Running our calculation up to 25,000 periods 
(or one million syzygies) we see that the noise level slowly rises in time, reaching the value 
$5 \cdot 10^{-6}$ towards the end of our computation. 
This is comparable with the cummulative numerical error in this calculation. 

\subsection{The $k=7,11,14$ slaloms}

Other solutions (denoted by $O\#$) shown in Table \ref{tab:1} correspond to 
%four up to scaling 
different slalom powers $k$. The first one, denoted by $O1$, belongs to the same slalom 
power $k=7$ as the new choreography solution, but it is not a choreography.
The second one $O2$ is the same solution as the first one $O1$, up to a rotation and scaling
of space-time. The two slalom power $k=11$ solutions $O3$ and 
$O4$ are also identical. The solution $O5$ has slalom power $k=14$. 

\subsection{Scaling of $k=17$ slaloms}

\begin{figure}[h!]
\centerline{\includegraphics[width=0.99\columnwidth,,keepaspectratio]{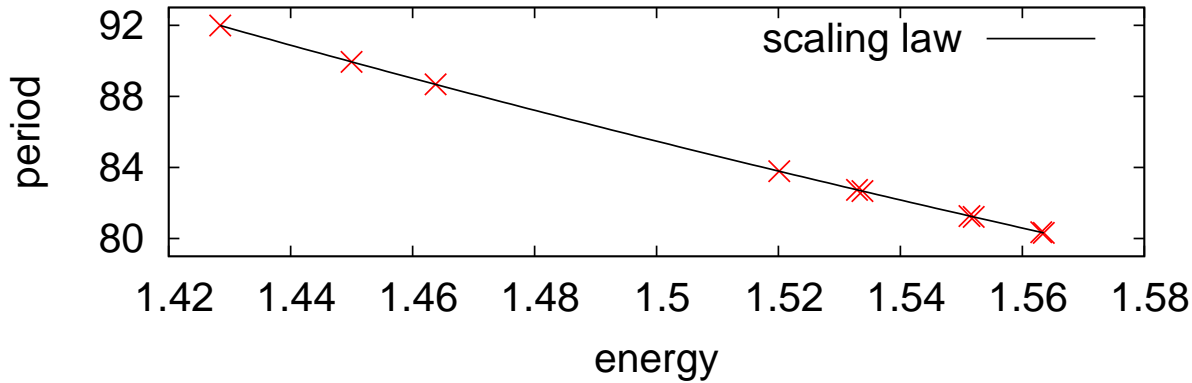}}
\caption{Red crosses: period $T$ vs. absolute value of energy $\varepsilon$ %dependence 
for slalom orbits with power $k=17$. Solid line: fit according to the scaling law 
$T \sim \varepsilon^{-3/2}$ (for details, see text).} 
\label{fFIGfit}
\end{figure}

All other solutions ($O6$-$O15$) have the slalom power $k=17$,
but demand special attention to determine if they are identical.
In Figure \ref{fFIGfit} we show the period $T$ as a function of the absolute value of the 
energy $\varepsilon = |E|$ %dependence 
for these orbits. The fitting parameter $A$ for the fit shown in Figure \ref{fFIGfit} with the
scaling law $T = A \varepsilon^{-3/2}$ is: $A=157.036 \pm 0.0007235$, with the reduced 
$\chi$-square value of $\chi^2_{red}=1.51197 \times 10^{-6}$. Such good agreement with the scaling 
law indicates that all these solutions may correspond to the same orbit. \

A careful analysis of passages through Euler points shows, however, that there are seven different 
$k=17$ slaloms. Solutions $O6$ and $O7$ correspond to same orbit. This is also the case with 
$O10$ and $O11$, and with $O14$ and $O15$. 
On the shape sphere each of these pairs of solutions has exactly the same trajectory, whereas their 
real-space trajectories are related by scaling transformations.

\subsection{Geometric-algebraic classification of results}

All solutions presented here fall into three new algebraic-geometric classes 
according to the classification scheme defined in Ref. \cite{Suvakov2013}. 
%According to geometric symmetry on the shape sphere, t
There were two different geometric classes defined in \cite{Suvakov2013}: 
(I) those with (two) reflection symmetries about two orthogonal axes --- the 
equator and the zeroth meridian passing through the "far" collision point; and 
(II) those with a (single) central reflection symmetry about one point --- 
the intersection of the equator and the aforementioned zeroth meridian.
Here, we have found two additional classes: (III) those with reflection symmetries 
about only one axis --- the equator; and (IV) those without any geometric symmetry 
on the shape sphere.

Similarly, in Ref. \cite{Suvakov2013} orbits were divided according to the
algebraic exchange symmetries of (conjugacy classes of) their free group elements: 
(A) with free group elements that are symmetric under 
$a \leftrightarrow A$ and $b \leftrightarrow B$, 
(B) with free group elements symmetric under $a \leftrightarrow b$ and 
$A \leftrightarrow B$, and (C) with free group elements that are not symmetric 
under either of the two symmetries (A) or (B).

We have observed empirically in Ref. \cite{Suvakov2013} that, for all orbits presented
there, the algebraic symmetry class (A) corresponds to the geometric class (I), and that the algebraic
class (C) corresponds to the geometric class (II), whereas the algebraic class (B) may  fall into
either of the two geometric classes. 

However, our new choreography solution does not obey this rule. It belongs to the 
algebraic symmetry class (A), but corresponds to the geometric class (II). This defines a new
geometric-algebric class II.A. The remaining solutions (slaloms) presented here define two 
additional classes: III.A and IV.A (see the last column in table \ref{tab:1}).

\section{Conclusions}
\label{conclusion}

In conclusion, as the result of numerical studies, we have found 11 new three-body solutions that 
can be described as slaloms with powers $k=7,11,14,17$. 
One of these solutions ($NC1=NC2$), with $k=7$, is a stable choreography. This particular orbit ought 
to be of special interest to mathematicians interested in rigorous existence proofs of three-body solutions. 
Other new non-choreographic orbits hold the same general interest as the solutions found in 
Ref. \cite{Suvakov2013}.

\begin{acknowledgements}
The author would like to thank V. Dmitra\v{s}inovi\'{c} for his help in the early stages of this 
work and with the write-up of this paper, and to Prof. Carles Sim\'{o} for providing
information about his solutions.
This work was supported by the Serbian Ministry of Education, Science and Technological 
Development under grant numbers OI 171037 and III 41011. The computing cluster Zefram 
(zefram.ipb.ac.rs)
%named after Dr. Cochrane, warp drive inventor in the Star Trek universe, 
has been extensively used for calculations. 
\end{acknowledgements}

\end{document}